\documentstyle{europhys}

\def\And{{\rm and\ }}

\newif\ifboo \boofalse

\def\Review#1{\boofalse{\it #1},}
\def\Name#1{{\sc #1},}
\def\Vol#1{\ifboo Vol. {\bf #1}\else{\bf #1}\fi}
\def\Year#1{\ifboo #1\else(#1)\fi}

\def\Page#1{\ifboo {\rm p. #1}\else{\rm #1}\fi}

\input epsf.sty
\begin{document}
\euro{}{}{}{}
\Date{}
\shorttitle{Z. \'A. N\'emeth \And J.-L. Pichard: Partially melted 
Wigner molecule}  

\title{Ground state of a partially melted Wigner molecule} 

\author{Zolt\'an \'Ad\'am N\'emeth$^{(a,b)}$  
\And Jean-Louis Pichard$^{(a)}$} 

\institute{
(a) CEA, Service de Physique de l'Etat Condens\'e, \\
Centre d'Etudes de Saclay, F-91191 Gif-sur-Yvette Cedex, France \\  
(b) E\"otv\"os University, Departement of Physics of Complex Systems, \\
H-1117 Budapest, P\'azm\'any P\'eter s\'et\'any 1/A, Hungary
}
\rec{}{}
\pacs{
\Pacs{71}{10-w} {Theories and models for many-electron systems}
\Pacs{71}{27+a} {Strongly correlated electron systems} 
\Pacs{73}{20.Qt} {Electron solids} 
}

\maketitle

\begin{abstract} 

 We consider three spinless fermions free to move on $2d$ square lattice 
with periodic boundary conditions and interacting via a $U/r$ Coulomb 
repulsion.  When the Coulomb energy to kinetic energy ratio $r_s$ is 
large, a rigid Wigner molecule is formed. As $r_s$ decreases, we show 
that melting proceeds via an intermediate regime where a floppy two 
particle molecule coexists with a partially delocalized particle. 
A simple ansatz is given to describe the ground state of this mesoscopic 
solid-liquid regime.  

\end{abstract} 
 
%
%
 Ordered arrays of charged particles with long range Coulomb repulsion 
have been a continuous subject of interest in various branches of physics, 
including colloidal suspensions, ion rings, atomic-ion Wigner crystals, 
quantum computers, biophysics, plasmas, electrons deposited on liquid Helium 
surfaces, charges created in semiconductor or organic field effect 
devices. These arrays can melt, exhibiting a transition from collective 
towards independent-particle motion, either as a function of the 
temperature (classical melting) or as a function of the charge density 
(quantum melting) at very low temperature. In principle, the quantum 
melting can be observed using electrons trapped in quantum 
dots \cite{ashoori,tarucha} or cooled ions confined in radio frequency 
traps \cite{walther}. Very often, a parabolic  confinement is imposed. 
When the confinement is weak and at a sufficiently low temperature, 
the Coulomb repulsion dominates the kinetic energy, the charges are 
ordered and a Wigner molecule is formed. If the confinement becomes 
stronger, the kinetic energy dominates the Coulomb repulsion, the 
molecule melts and one gets a Fermi system of weakly interacting particles. 
In a parabolic $2d$ trap, the molecule consists of well-separated shells. 
Both for the classical melting \cite{peeters} (achieved by increasing the 
temperature for a given trap) and for the quantum melting \cite{filinov} 
(achieved at zero temperature by reducing the size of the trap), it has been 
observed that melting proceeds in two stages: first neighboring shells 
may rotate relative to each other while retaining their internal 
order, second the shell broadening leads to radial melting. Wigner 
quantum crystallization in $2d$ electron dots is characterized by 
two distinct - radial and orientational - ordering transitions. However, 
a parabolic trap does not yield a uniform charge density, the low density 
shells at the edges could order before the high density part in the bulk, 
and this two stage melting could be an artifact due to the interplay 
between surface and bulk orderings. It is therefore interesting to study 
if a multi stage melting persists in a  system of uniform charge density, 
for instance when the charges are confined on a $2d$ torus. One has then 
to take into account the translational symmetry of a $2d$ torus instead 
of the rotational symmetry of a parabolic trap.

%
%

 Another important issue can be mentioned, assuming that insights 
gained through investigations limited to small systems provide the 
foundations for understanding larger systems. Long ago, it was suggested 
by Andreev and Lifshitz \cite{andreev-lifshitz} that low temperature 
localized defects change into excitations that move practically freely 
through a crystal. As a result, the number of sites of a quantum crystal 
may be smaller than the total number of particles present in the system 
for intermediate couplings, such a crystal being neither a solid, nor a 
liquid. Two kinds of motion are possible in it; one possesses the 
properties of motion in an elastic solid, the second possesses the 
properties of motion in a liquid. 

%
%

 An intermediate regime of melting was detected using $N=4$ spinless 
fermions in a $L\times L$ lattice, and it was observed 
\cite{katomeris,windsor} that a combination of a few plane waves and 
site orbitals was able to describe it, suggesting a liquid-solid 
regime consistent either with a scenario \`a la Andreev-Lifshitz, or 
with a possible quantum liquid crystal regime \cite{bwp2,okf}. This 
intermediate regime was shifted to lower ratios $r_s$ when site disorder 
was included. Many signatures of a novel ground state (GS) were observed 
for intermediate couplings, considering the structure \cite{bwp1} of 
the persistent currents when the torus was pierced by an Aharonov-Bohm 
flux, the statistics \cite{bwp2} of its low energy excitations, 
the failure \cite{windsor} of the Hartree Fock approximation to 
describe the persistent currents above a first ratio $r_s^F$, the 
suppression of the same currents above a higher ratio $r_s^W$, the 
GS magnetization \cite{selva} when the spin degrees of freedom where 
included. Moreover, it was noticed in Ref. \cite{bwp1} that the 
ratios $r_s^F <r_s < r_s^W$ where the novel mesoscopic GS was 
observed were consistent with those where transport measurements 
using dilute electron gases in $2d$ field effect devices 
\cite{kravchenko,batlogg} indicate the puzzling possibility of a 
novel $2d$ metal. 

%
%

 The purpose of this work is to reveal the exact nature of the 
intermediate mesoscopic GS and to describe it with a simple 
wave function, considering $N=3$ spinless fermions with 
$U/r$ Coulomb repulsion in a $L \times L$ square lattice with 
periodic boundary conditions. Using the operators  $c_j^\dagger$ 
($c_j$) , $c_k$ ($c_k^\dagger$) which create (destroy) a spinless 
fermion either at the lattice site $j=(j_x,j_y)$ or in a plane wave 
state of momentum $k=(k_x,k_y)$ of this lattice, the Hamiltonian reads:
\begin{equation}
H=-t \sum_{\left<j,j'\right>}c_j^\dagger c_{j'} + {U\over2} \sum_{j, j' 
\atop j\neq j'}{n_j  n_{j'}\over d_{j,j'}}= 
\sum_k \epsilon_k c_k^\dagger c_k +
\sum_{k,k',q} U(q) c_{k+q}^\dagger c_{k'-q}^\dagger c_{k'} c_k. 
\label{Hamilton}
\end{equation}
$n_j=c_j^\dagger c_j$, $d_{j,j'}$ is the shortest distance between 
sites $j$ and $j'$, $\epsilon_k=-2t (\cos k_x +\cos k_y)$  and 
$U(q)=U/(2 L^2) \sum_{j} \cos (qj)/d_{0,j}$. The Coulomb energy to 
kinetic energy ratio $r_s= U/(2t\sqrt{\pi n_e})$ for a filling factor 
$n_e=N/L^2$. The operators $c_k$ and $c_j$ are related by the usual 
Fourier transform:
\begin{equation} 
c_k= {\frac{1}{L}} \sum_j e^{- i (kj)} c_j. 
\end{equation}

 In the eigenbasis of the non interacting system (eigenvectors 
$c_{k_1}^\dagger c_{k_2}^\dagger c_{k_3}^\dagger \left|0\right>$, 
$\left|0\right>$ being the vacuum state), the Hamiltonian matrix is block 
diagonal, each block being characterized by the same conserved total 
momentum $K=\sum_{i=1}^3 k_i$. Moreover, only the non interacting states 
having in common one $k$ out of three can be coupled by the interaction 
inside a $K$ sub-block. Therefore, each $K$ sub-block is a sparse matrix 
which can be exactly diagonalized using the Lanczos algorithm when 
$L$ is small enough.

\begin{figure}
\begin{center}
\begin{minipage}{3.5cm}
{\centerline{\leavevmode \epsfxsize=3cm \epsffile{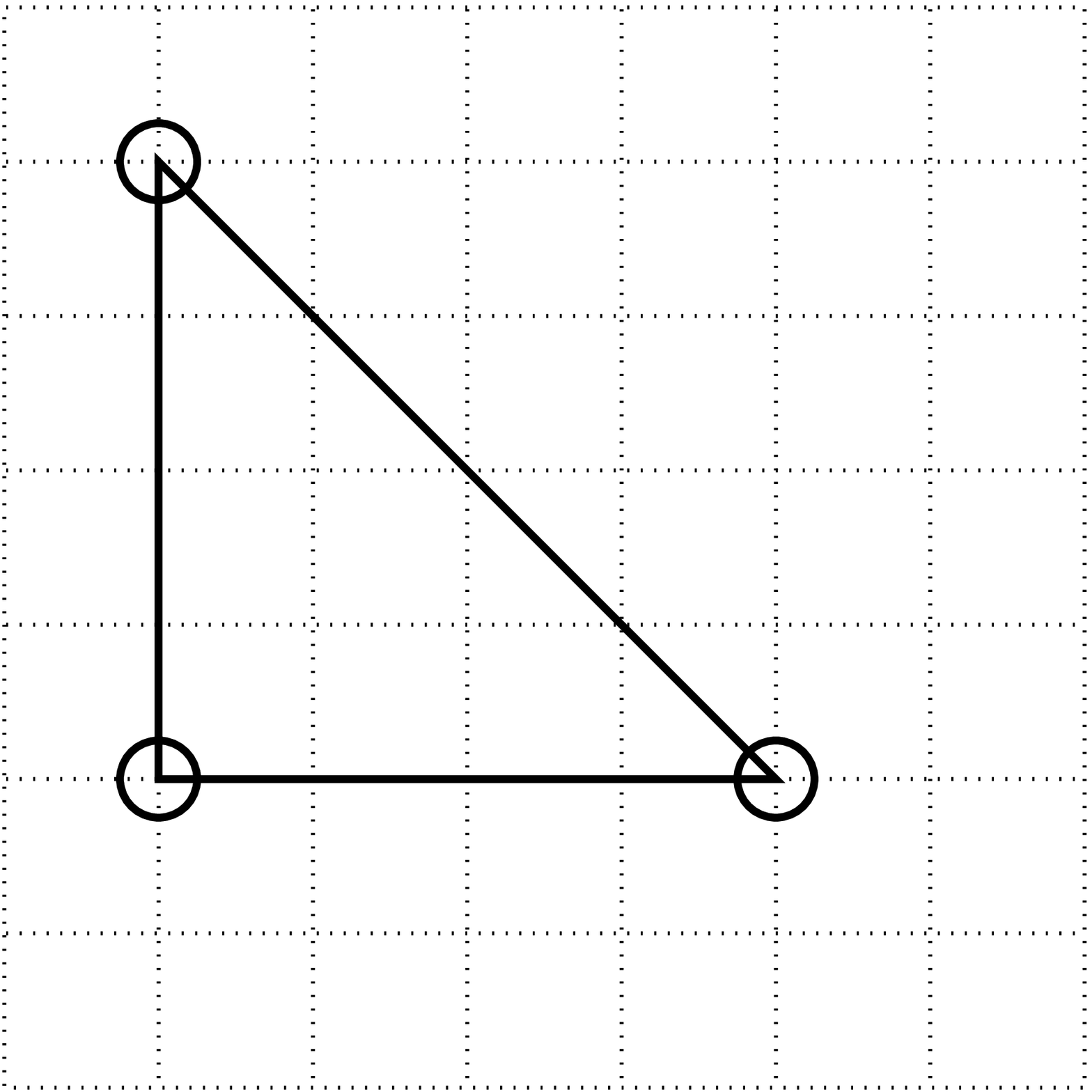}}}
\centerline{$K=({2 \pi \over 8},{2 \pi \over 8})$}
\end{minipage}
\begin{minipage}{10.5cm}
{\centerline{\leavevmode \epsfxsize=8.5cm \epsffile{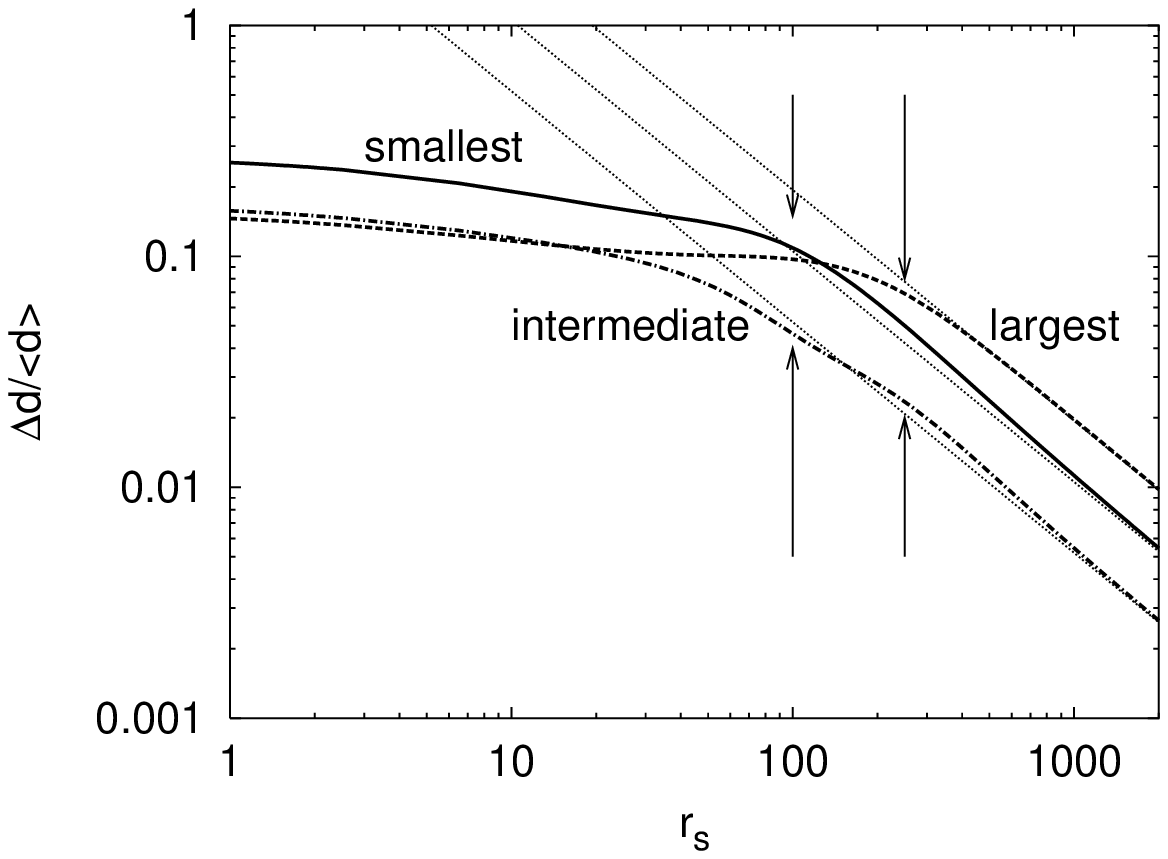}}}
\end{minipage}
\begin{minipage}{3.5cm}
{\centerline{\leavevmode \epsfxsize=3cm \epsffile{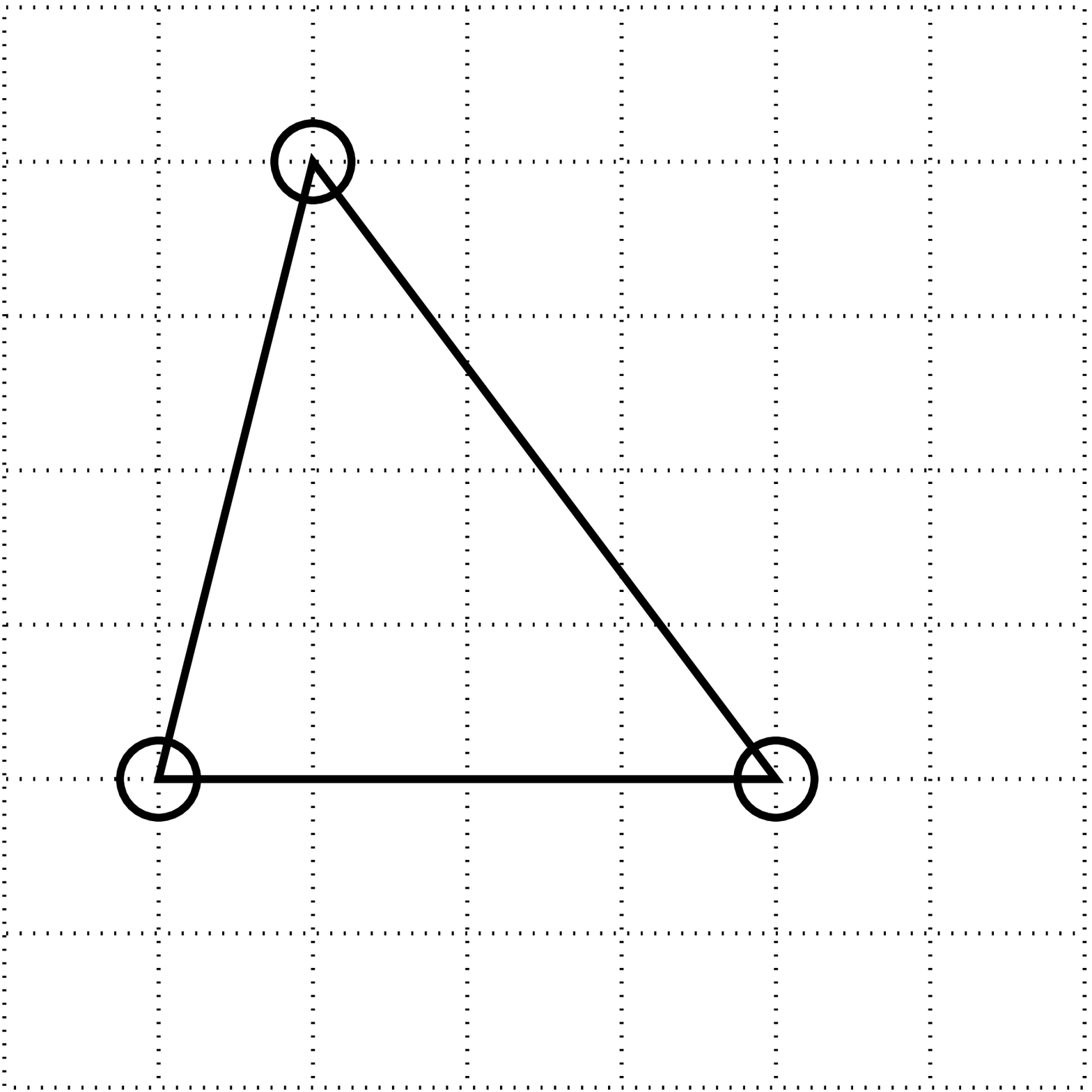}}}
\centerline{$K=(0,0)$}
\end{minipage}
\begin{minipage}{10.5cm}
{\centerline{\leavevmode \epsfxsize=8.5cm \epsffile{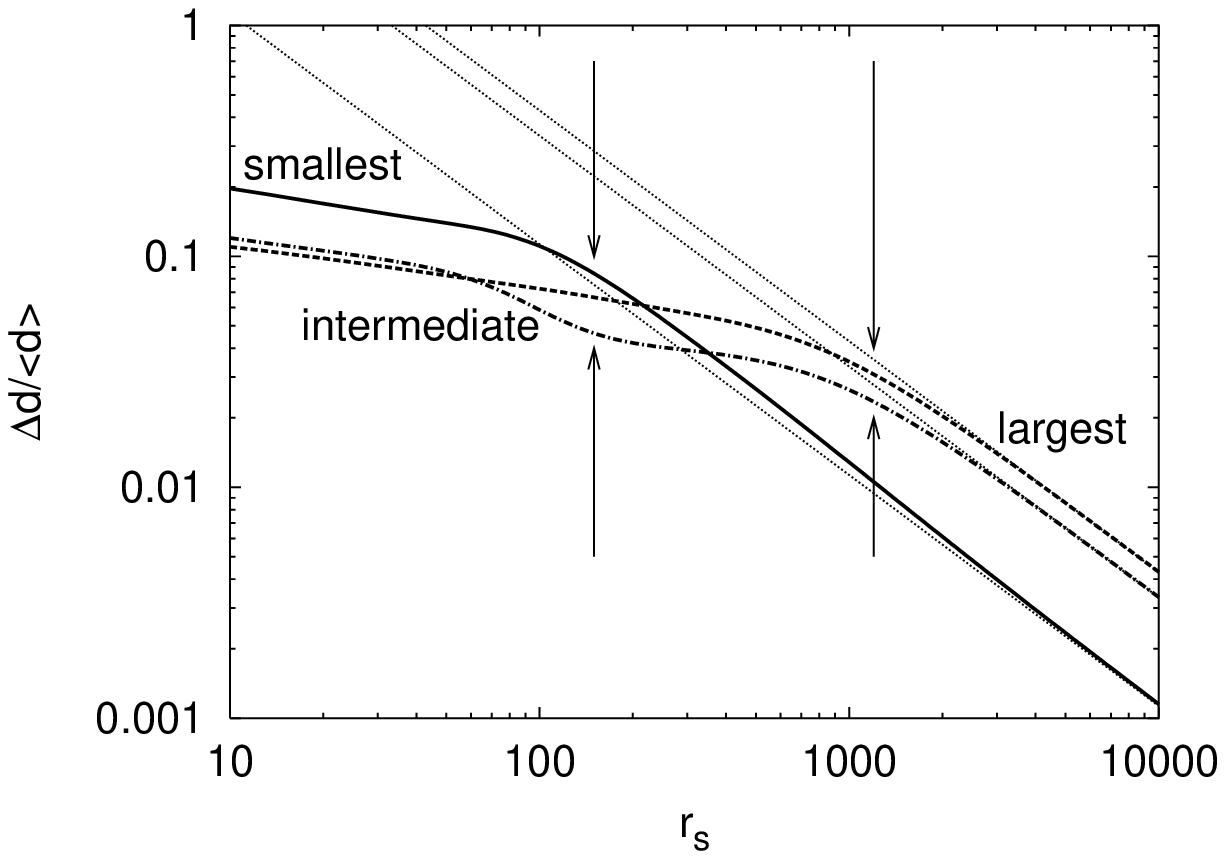}}}
\end{minipage}
\end{center}
\label{Fig1}
\caption{Left sides: Scheme of two low energy Wigner molecules 
in the limit $r_s \rightarrow \infty$ and of total momenta 
$K$ for $L=8$. Right sides: Corresponding relative fluctuations 
$\Delta d /\left<d\right>$ of the three particle spacings as a function 
of $r_s$. Smallest (continuous line), intermediate (dashed dot line) 
and largest (dashed line) spacings. The thin dotted lines give 
the $t/U$ perturbative behaviors. The arrows indicate the two crossover 
ratios $r_s^*$}.
\end{figure}

%
%

 In the large coupling limit ($r_s \rightarrow \infty$) the eigenstates 
correspond to rigid Wigner molecules. For a localized center of mass, 
the charges would be simply located on three lattice sites $a, b $ and $c$, 
the location of those sites minimizing the electrostatic energy. 
However translational invariance implies a delocalization of the center 
of mass in a plane wave state of momentum $K$, and the 
Wigner molecule wave functions become:
\begin{equation}
\left|\Psi\right>={1\over L}\sum_j e^{i (Kj)} c^\dagger_{j+a} 
c^\dagger_{j+b} c^\dagger_{j+c} \left|0\right>.
\label{order0}
\end{equation}
For a given shape $(a,b,c)$ of the three particle molecule, one has three 
well defined inter particle spacings $d_{min} \leq d_{int} \leq d_{max}$. 
The electrostatic energy becomes $ U(d_{min}^{-1}+d_{int}^{-1}
+d_{max}^{-1})$ while the kinetic energy $\propto t_{eff}$ (center 
of mass effective band width)  $ \rightarrow 0$ as 
$r_s \rightarrow \infty$.  For some arbitrary values of $K$ and for some 
molecular shapes of low electrostatic energy, we have decreased $r_s$ 
and followed the corresponding levels, ignoring possible 
level crossings with other levels of different $K$ and of different 
molecular shape. Two examples calculated using a $L=8$ lattice are 
shown in Fig. 1, revealing the generic scenario for the 
melting of a three particle Wigner molecule. If one considers the 
change of the relative fluctuations $\Delta d /\left<d\right>$ as 
one increases $r_s$, 
one can see both for $d_{min}$ and $d_{max}$ a crossover 
from a weak coupling behavior where the fluctuations are large 
(the $d$ are not well defined) towards a large coupling behavior where 
the fluctuations become negligible (the $d$ become well defined). The 
weak (strong) coupling limits can be described using $U/t$ ($t/U$) 
perturbative expansions. For instance, the large coupling behavior 
of the Wigner molecule is given at first order of a  $t/U$ expansion 
by:  
\begin{equation}
\left|\Psi (1) \right>=\left|\Psi\right>+
\sum_{\alpha=1 \atop \alpha\neq 0}^{12}{t\over 
\Delta E_\alpha}\left|\Psi_{\alpha}\right>, 
\label{vect-prop}
\end{equation}
where the $ \left|\Psi_{\alpha}\right>$ label the Wigner molecules 
of same $K$ obtained by moving one of the sites $a,b,c$ of 
$\left|\Psi\right>$  by one lattice spacing, 
$\Delta E_\alpha \propto U$ being the 
corresponding changes of electrostatic energy. This gives the $t/U$ 
decays of the three $\Delta d/\left<d\right>$ indicated in Fig. 1. 

The main point to notice is the clear separation 
between the crossover ratios $r_s^*$ (indicated by the arrows in 
Fig. 1) characterizing $d_{min}$ and $d_{max}$. As one can see 
in the data, there are relatively large intervals of intermediate 
couplings where $d_{min}$ is well defined while $d_{max}$ is not, 
giving rise to an intermediate behavior for $d_{int}$. This tells us 
that the generic melting of a three particle molecule proceeds also 
in two stages, if one considers a $2d$ system of uniform density, 
as it had been shown using $2d$ parabolic traps. The intermediate 
regime of melting consists of a floppy two particle molecule co-existing with 
a third  delocalized particle.

%
%

 We now study the ground states (GSs) of the three body 
problem. The GS momenta and degeneracies depend on $L$, as 
the possible existence of GS level crossings. For simplicity, 
let us consider the case where $L$ is even. At $U=0$, one has a 
sixfold GS degeneracy which is partially removed by an infinitesimal 
$U$, the energy of a set of four states with momenta $K= 
(\pm 2\pi/L,\pm 2\pi/L)$ and $ K= (\pm 2\pi/L,\mp 2\pi/L)$ becoming 
different to those of the two $K=0$ states. 
Using a $U/t$ expansion, one finds that the GSs remain in the first 
set for $L \leq 6$ while they go into the second set for $L \geq 8$. 
At $t=0$, the low energy Wigner molecules are $L^2$ triangles 
($d_{min}=d_{int}=L/2, d_{max} = L/\sqrt 2$) having $L^2/4$ different 
locations of their centers of mass and $4$ different orientations. 
This $L^2$ degeneracy is removed when one turns on $t$. The energies 
$E_0(K)$ of the $L^2$ first levels are given when $t/U \rightarrow 0$ 
by:
\begin{equation} 
E_0(K) \approx A -2 t_{eff} \left (\cos K_x+\cos K_y \right ) + 
2r_{eff} \left ( \cos (K_x L/2) + \cos (K_y L/2)  \right),
\end{equation} 
where $A$ is a K-independent energy, $ t_{eff} \propto t (t/U)^{N-1}$ 
is the effective center of mass band width while $r_{eff} \propto 
t(t/U)^{L/2-1}$ is the effective band width coming from single 
particle motions which couple triangles of same center of mass but 
of different orientations  ($L/2$ one particle hops). For $L \geq 8$, 
$t_{eff} >>  r_{eff} $, one has a non degenerate $K=0$ GS when 
$t/U \rightarrow \infty$, and consequently a GS level crossing 
as $r_s$ increases inside the $K=0$ subspace between the two weak 
coupling GSs and the single large coupling GS. For $L=6$, $r_{eff}$ 
and $t_{eff}$ are both $ \propto t^3/U^2$, the GSs keep as 
$r_s$ varies their momenta $K= (\pm 2\pi/L,\pm 2\pi/L)$ and 
$ K= (\pm 2\pi/L,\mp 2\pi/L)$ (fourfold degeneracy) and 
no GS level crossing occurs. Hereafter we study the $L=6$ GS of 
momentum $K=(2\pi/6,2\pi/6) $. This allows us to avoid the 
complications coming from the GS level crossing for $L \geq 8$. 
However, as shown by the previous examples, our results will be 
relevant to generically describe the multi stage quantum melting 
of a $N=3$ low energy Wigner molecule if one continuously follows 
a given level from large couplings towards weak couplings. 

%
%

\begin{figure}
\begin{center}
\begin{minipage}{3.5cm}
{\centerline{\leavevmode \epsfxsize=3cm \epsffile{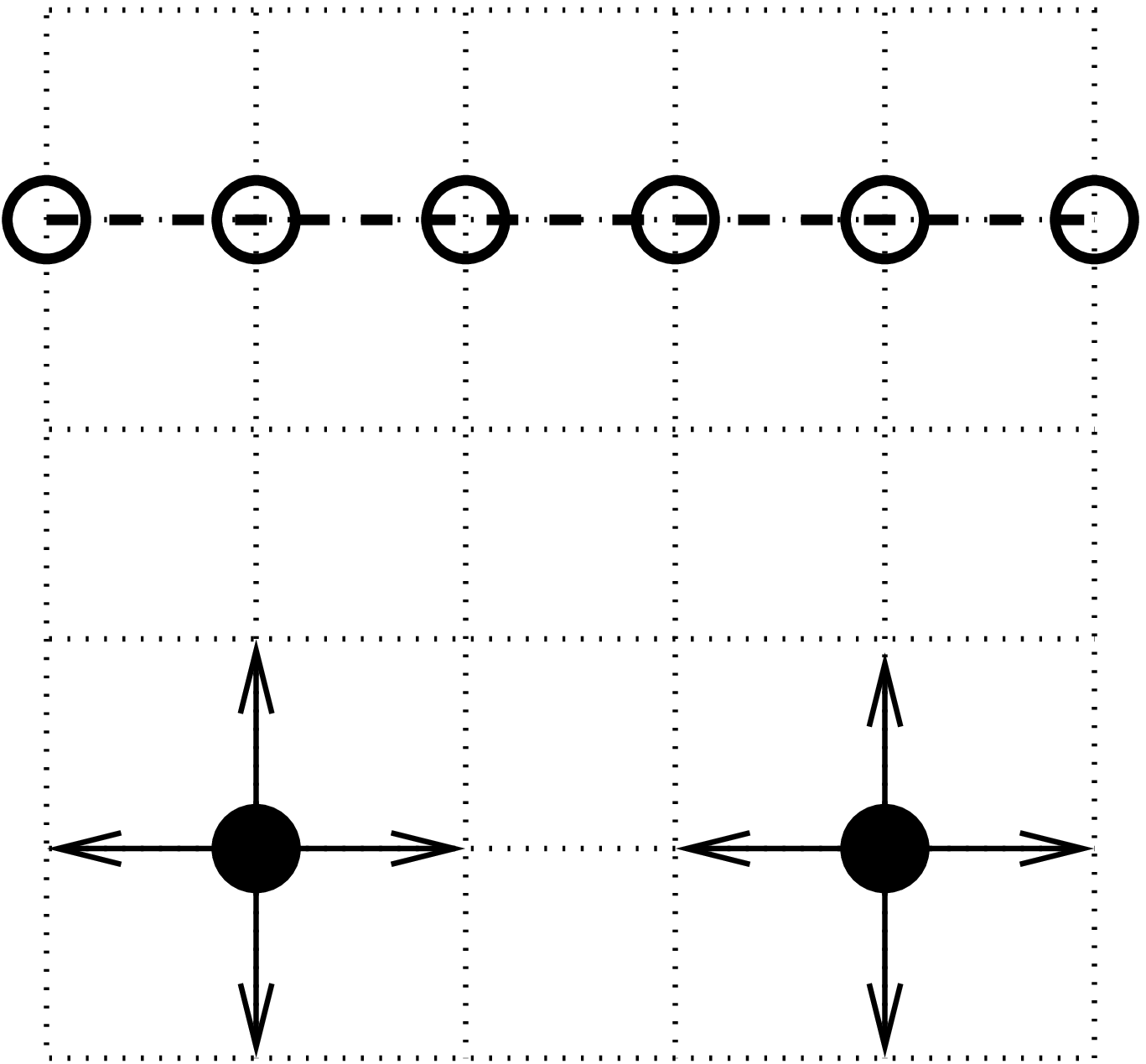}}}
\centerline{$K=({2\pi\over6},{2\pi\over6})$}
\end{minipage}
\begin{minipage}{10.5cm}
{\centerline{\leavevmode \epsfxsize=8.5cm \epsffile{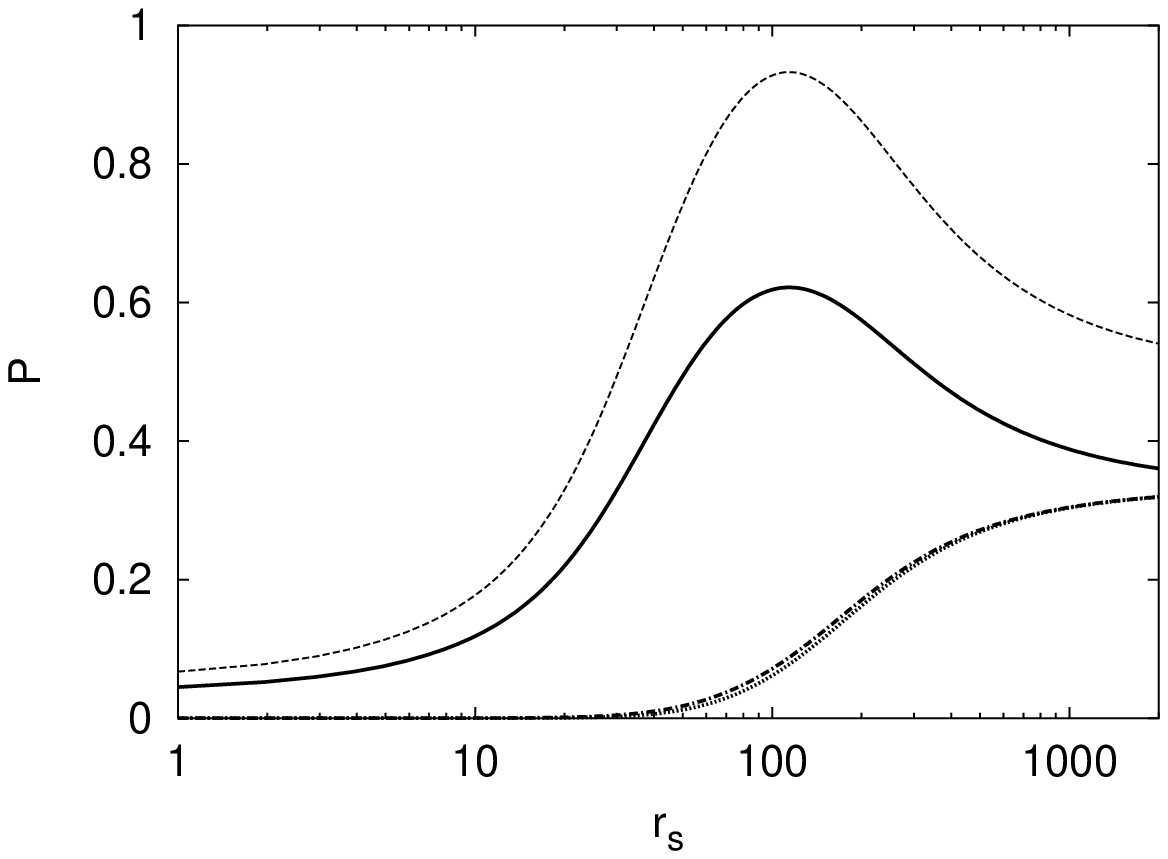}}}
\end{minipage}
\end{center}
\caption{Left figure: Scheme of a x-oriented partially melted triangular 
molecule (x-PMTM) for $L=6$. The arrows give configurations allowed by a 
partial $t/U$ expansion of the 2PWM. Right figure: GS projections 
$P_x (K,k_x=0)$ (solid line) $P_x (K, k_x=2\pi/3)$ (dashed-dotted line), 
$P_x (K,k_x=-2\pi/3)$ (dotted line) over x-PMTMs  of momentum $k_x$. 
$P(K,k_x=0,k_y=0)$ (upper thin dashed line) gives the total projection 
over the PMTM of momenta $k_x=0$ and $k_y=0$. Total momentum 
$K=(2\pi/6,2\pi/6)$.
}
\label{Fig.2}
\end{figure}

\begin{figure}
\begin{center}
\begin{minipage}{7cm}
{\centerline{\leavevmode \epsfxsize=7cm \epsffile{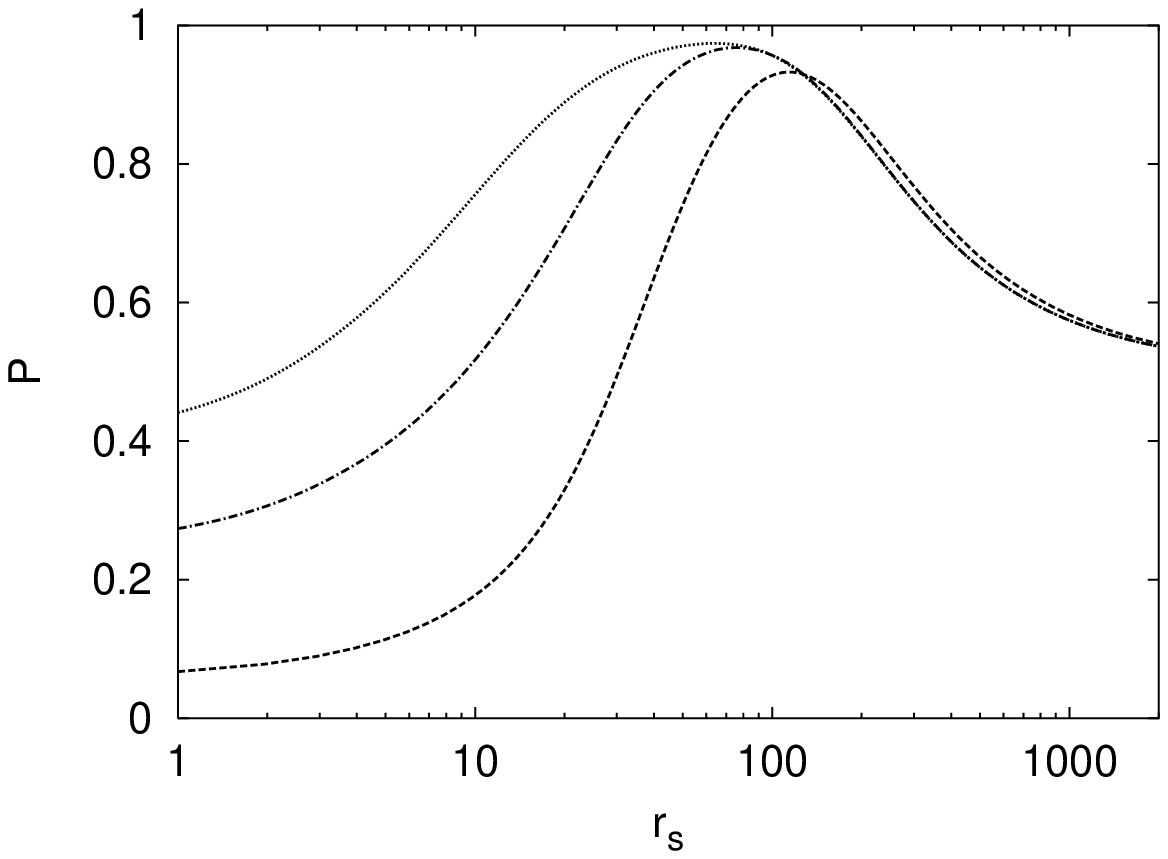}}}
\end{minipage}
\begin{minipage}{7cm}
{\centerline{\leavevmode \epsfxsize=7cm \epsffile{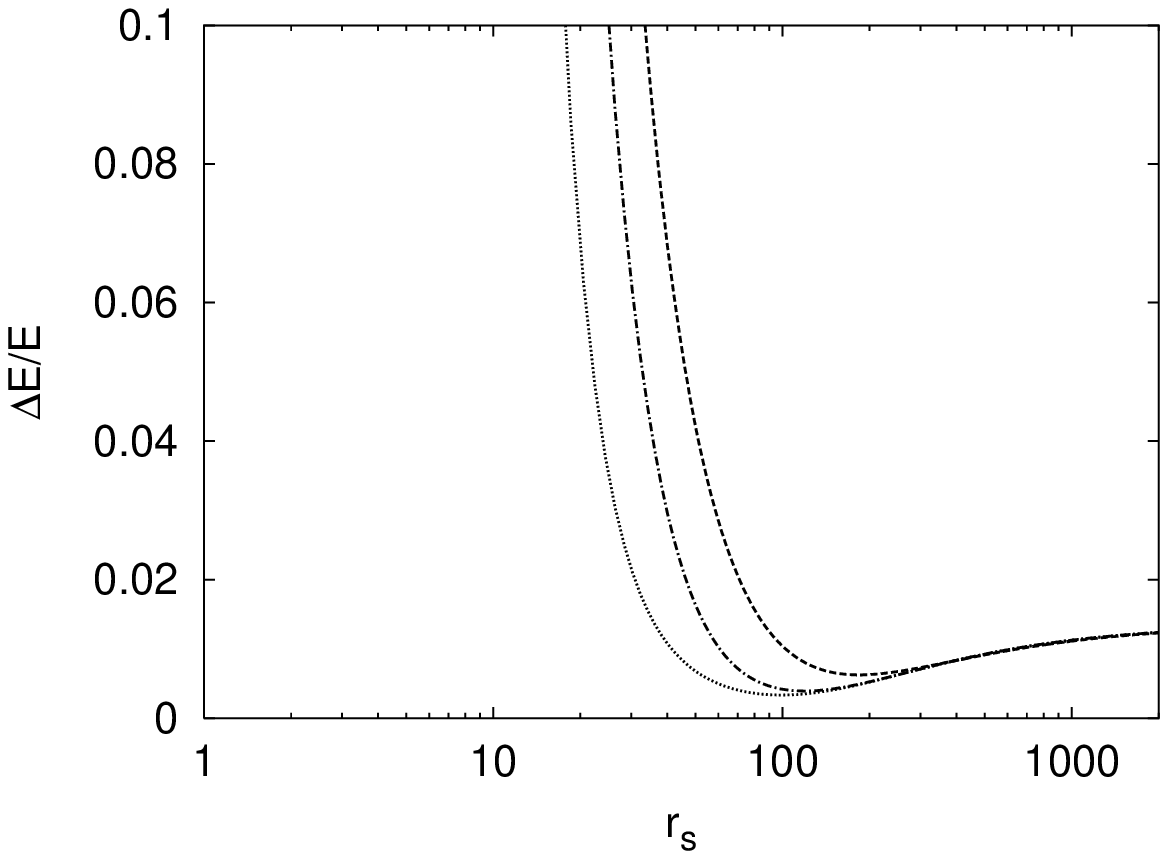}}}
\end{minipage}
\begin{minipage}{7cm}
{\centerline{\leavevmode \epsfxsize=7cm \epsffile{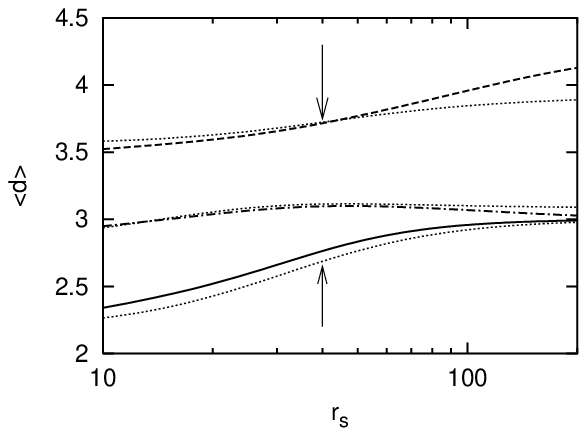}}}
\end{minipage}
\begin{minipage}{7cm}
{\centerline{\leavevmode \epsfxsize=7cm \epsffile{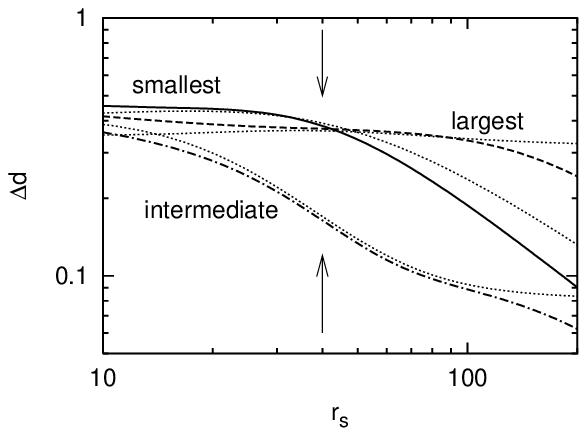}}}
\end{minipage}
\end{center}
\caption{
Upper left: GS projections over PMTM ansatz with $k_x$ and 
$k_y=0$; Upper right: Relative errors $\Delta E/ E$; The bare 
ansatz behaviors, corrected by a first order or a second order 
$t/U$ expansion of 2PWM  are given by dashed lines, 
dashed-dotted lines and the dotted lines respectively;  
Lower left: Average inter-particle spacings $\left<d\right>$; 
Lower right: Fluctuations $\Delta d$ of the inter-particle spacings 
as a function of $r_s$. Exact behaviors (thick lines), 
and ansatz behaviors corrected by a second order $t/U$ expansion 
of 2PWM. $K=(2\pi /6,2\pi /6)$.
}
\label{Fig.3}
\end{figure}

 For $L=6$, the degeneracy of the $L^2$ triangular molecules 
is broken when one turns on $t$. A $t/U$ expansion gives for 
the $L^2=36$ low energy levels
\begin{equation}
E_0(K)=A_2 - 2 t_3 \left(\cos 2\pi K_x+\cos 2\pi K_y \right) + 2 r_3
\left(\cos 2\pi 3 K_x+\cos 2\pi 3 K_y  \right) + 0({\frac{t^4}{U^3}}). 
\label{etrinagl}.
\end{equation}
$A_2 = 0.9023 U - 208.9 (t^2/U)$, $ t_3=1000  (t^3/U^2)$ and 
$r_3=1660 (t^3/U^2)$. This $t/U$ expansion makes sense when both 
$d_{min}$ and $d_{max}$ are well defined, with small relative 
fluctuations of order $t/U$. This means $r_s \geq 200$ (see 
Fig. \ref{Fig.3}). To describe lower $r_s$ ($ 40 < r_s < 200$), 
where $d_{min} \approx 3$ is well defined, while $d_{max}$ has still 
large fluctuations, we propose a simple ansatz based on the concept of 
partially melted triangular molecules (PMTMs). A x-oriented PMTM 
(x-PMTM) is a rigid two particle Wigner molecule (2PWM) with 
$d_{min} = L/2$ combined with a third particle free to 
move with a wave vector $k_{x}$ parallel to the 2PWM at a distance 
$L/2$, as sketched in Fig. \ref{Fig.2}. 
The x-PMTM wave function of momentum $K$ reads: 
\begin{equation}
\left|\Psi_x (K,k_{x})\right>= {1\over6 \sqrt{2}} \sum_j e^{ i (K_x-k_{3x}) 
j_x + K_y j_y} c_{j+a}^\dagger c_{j+b}^\dagger c_{k_{x},j_y+c_y}^\dagger 
\left|0\right>,
\end{equation}
where $a=(0,0), b=(3,0), c=(0,3)$, and 
\begin{equation}
c_{k_{3x},j_y+c_y}^\dagger 
= {1 \over \sqrt 6} \sum_{j_{x'}} e^{i k_{x} j_{x'}} 
c_{j_{x'},j_y+c_y}^\dagger.
\end{equation}
$(K_x-k_{x})\cdot 6 /(2\pi)$ must be odd, which leads to 
$k_{x} =0, \pm 2\pi/3$ for $K=(2\pi/6,2\pi/6)$. The y-oriented 
PMTM wave function $\left|\Psi_y(K,k_{y})\right>$ is defined in a 
similar way. 
The final PMTM ansatz of momentum $K$ is a normalized combination 
of the x and y-PMTMs, which reads:
\begin{equation}
\left |\Psi(K,k_x,k_y) \right>={\left|\Psi_x(K,k_{x})\right>-
\left|\Psi_y(K,k_{y})
\right>\over \sqrt{ 2-2\left< \Psi_x(K,k_{3x})\mid \Psi_y(K,k_{3y})\right>}}
=\sqrt{3\over8}\left(\left|\Psi_x(K,k_{x})\right>-\left|\Psi_y(K,k_{y})
\right>\right),
\label{projtot}
\end{equation}
and the constraint $k_{x}=k_{y}$ makes it invariant under $x-y$ 
permutation.

 In Fig. 2, the three values $P_x (K,k_x)=|\left <\Psi_0 (K) \mid 
\Psi_x(K,k_{x})\right>|^2$ taken by the projections of the exact 
GS  $\left|\Psi_0(K)\right>$  over the x-PMTMs $\left |\Psi_x(K,k_{x}) 
\right>$ are given as a function of $r_s$, together with the GS projection 
$P (K,k_x,k_y)=|\left <\Psi_0 (K) \mid \Psi (K,k_x,k_y)\right>|^2$ 
over the PMTM of momenta $(K,k_x=0,k_y=0)$. 
Following the three projections over the x-PMTMs 
of different wave vector $k_x$, one can see how the third particle 
gets progressively localized in the x-direction as $r_s$ increases, 
the rigid three particle triangular molecule corresponding to 
$P_x (K,k_x)= 1/3$ for the three possible $k_{x}$. 
$P (K,k_x=0,k_y=0) \approx 93 \% $ at $r_s \approx 100$. 
However, only the PMTMs with $k_{x}=k_{y}=0$ contribute  
when $r_s \leq 50$. For those values of $r_s$, the third particle 
is fully delocalized in the direction parallel to the oriented PMWMs.
However, it is likely that the PMTM ansatz overestimates the rigidity 
of the remaining 2PWM when $r_s$ becomes smaller. 
This can be partly fixed using a $t/U$ expansion for the 2PWM 
(as sketched in Fig. 2 left) and keeping the third particle in its 
delocalized plane wave state with $k_x (k_y)=0$.

 The improvements coming from this partial $t/U$ expansion 
of the PMTM ansatz are given in Fig. 3, where one can see 
the behaviors of the bare ansatz, of the ansatz corrected to first 
order and to second order of the $t/U$ expansion of the 2PWM.
In the upper figures, the GS projections and the relative errors 
$\Delta E(p) /E$ are shown, $E$ denoting 
the  exact GS energy, $\Delta E (p) = E_A(p)-E$, $E_A(p)$ being the 
ansatz energy at the $p^{th}$ order of the partial $t/U$ expansion. 
Not only the GS description is improved, but lower values of $r_s$ 
can now be reached. In the lower figures, the three GS interparticle 
spacings $d_{min}$ $d_{int}$ and $d_{max}$ are given, and compared to 
the corresponding values of the ansatz, after a second order $t/U$ 
expansion of the 2PWM. As underlined by the arrows, both the averages 
and the variances are now well described for $r_s \approx 40$. 

However, let us underline that the mesoscopic melting process is 
not yet achieved at $r_s \approx 40$. From a study of the weak 
coupling limit, we have obtained precursor behaviors of the formation 
of the Wigner molecule at smaller $r_s$. For instance, certain GS 
projections over low energy non interacting states which are close 
in energy to the non interacting GS, but orthogonal to the large 
coupling Wigner molecule begin to decay when $r_s > 5$. Therefore, 
the PMTM ansatz, even improved by a $t/U$ expansion of the 2PWM fails to 
describe this precursor regime ($ 5 < r_s <30$) where a floppy 2PWM 
takes place, but is not rigid enough to be described by a simple 
$t/U$ expansion. 

%

 In summary, we have shown that the quantum melting of a three particle 
Wigner molecule confined on a $2d$ torus proceeds via an intermediate 
regime which can be described by the simple concept of a partially 
melted Wigner molecule, built of a delocalized particle and of a 
floppy 2PWM. This is in agreement with the general multi stage 
picture of mesoscopic quantum melting given by other works using $2d$ 
parabolic traps. At a mesoscopic scale, this gives a simple  
illustration of the quantum crystal with $k=0$ defectons conjectured 
by Andreev and Lifshitz. Notably, one can see that the 
number of Wigner lattice sites is smaller than the total number of 
charges. This shows that the multi stage melting is not a mesoscopic 
surface effect and suggests that dilute $2d$ electron gases of 
intermediate $r_s$ could be more complicated than usually assumed. 

%
%
This work was supported by the EU program ``Nanoscale dynamics, coherence 
and computation'' 
and the Hungarian Science Foundation OTKA TO25866 and TO34832.

\end{document}